\documentstyle[12pt]{article}

\input epsfig.sty

\def\simge{\mathrel{%
   \rlap{\raise 0.511ex \hbox{$>$}}{\lower 0.511ex \hbox{$\sim$}}}}
\def\simle{\mathrel{
   \rlap{\raise 0.511ex \hbox{$<$}}{\lower 0.511ex \hbox{$\sim$}}}}

\newcommand{\beq}{\begin{equation}}
\newcommand{\eeq}{\end{equation}}

\newcommand{\al}{\alpha}
\newcommand{\be}{\beta}
\newcommand{\bl}{\bigl\langle}
\newcommand{\br}{\bigr\rangle}

\newcommand{\de}{\delta}
\newcommand{\di}{\displaystyle}

\newcommand{\la}{\lambda}

\newcommand{\si}{\sigma}

\newcommand{\ve}{\varepsilon}
\newcommand{\vk}{{\bf k}}
\newcommand{\vp}{\varphi}

\newcommand{\prd}{Phys. Rev. D}

\newcommand{\pr}{Phys. Rep. }

\newcommand{\plb}{Phys. Lett. B}

\title{Invariant dynamics of scalar perturbations
\protect\\
  of inflanton and gravitational fields.  }
\author{O.Lalakulich${}^\dagger$, L. Marochnik${}^\ddagger$,
      G. Vereshkov${}^\dagger$}
\date{${}^\dagger$ Research Institute of Physics,
Rostov State University, Russia
\protect \\
${}^\ddagger$ Computer Sciences Corporation / Space
Telescope Science Institute, NASA, USA }

\begin{document}
\maketitle

\begin{abstract}
     A   gauge-independent,   invariant   theory   of   linear  scalar
perturbations  of inflation and gravitational fields has been created.
This invariant theory allows one to compare gauges used in the work of
other researchers and to find the unambiguous criteria to separate the
physical  and coordinate effects. It is shown, in particular, that the
so-called  longitudial gauge, commonly used when considering inflation
instability,  leads  to  a  fundamental  overestimation  of the effect
because of non-physical perturbations of the proper time in the frame
of  reference specified by this gauge (see main text for the numbers).
Back  reaction  theories  employing  this  sort  of gauge also involve
coordinate  effects.  A  comprehensive  review  of papers that use the
longitude  gauge  to  analyze  back  reaction effects and inflationary
instability  can  be  found in [1]. The invariant theory created here
shows  that  the  classical Lifshitz (1946) [2] gauge does not lead to
non-  physical  perturbations of the proper time. This is true because
in  his  pioneering  work,  Lifshitz  used  the  synchronous  frame of
reference, which enables to use one and same time for the description
of  both  the  background  and  perturbation evolution. This gauge can
therefore  be  used  to  analyze  the  inflation  regime  and the back
reaction  of perturbations on this regime properly. The first theory
of back  reaction on background of all types of perturbations (scalar,
vector and tensor) based on this gauge was published in 1975 [3]. More
recently  it  has  been  applied  to  the  inflation  regime  [4]. The
investigation  of  long-length  perturbations,  which characterize the
stability of the inflationary process, and quantum fluctuations, which
form  the  Harrison-Zel'dovich  spectrum  at  the end of inflation, is
performed  in the invariant form. The invariant theory proposed allows
one  to examine the effect of quantum fluctuations on the inflationary
stage  when the periodic regime changes to an aperiodic one. Numerical
examples  are  presented  in  the  main text. The theory also properly
describes  the  spectrum  reconstruction  during  the  epoch  when the
inflationary  stage  changes  to the Friedman one. The final effect of
this  process  will  be investigated in future space experiments. That
only the invariant theory must be used to analyze space experiments is
one of the conclusions of the present work.
\end{abstract}

\section{Introduction.}

The present paper proposes a new approach to the two main problems
of the theory of scalar perturbations in the system of inflanton and
gravitational fields.  These problems have been extensively discussed
in the literature.

The first problem is about the dynamics of the perturbations the
length of which is comparable with the Universe size at the beginning
of inflation. Two aspects of the problem are: (i) the stability
of the inflationary process; (ii) the back reaction of the long-length
perturbations on the expansion of the Universe in the past and in the
present epoch. (The latter is one of the approaches to the
problem of dark energy.)

The second problem is how the Harrison--Zel'dovich spectrum
was formed from the inflanton field vacuum fluctuations, the length of
which is much less than the Universe size at the beginning of the
inflation.  As it is known, the solution of this problem is the
basis for the modern theory of the large scale structure formation
in the Universe.

The conventional approach to solving the problems mentioned above
is based on the investigation of the scalar perturbations in the
longitudial gauge. The theory is formulated in terms of the
relativistic scalar potential $\Phi(\vec x,t)=\de g_{00}(\vec x,t)$;
all other observable values are expressed via $\Phi(\vec x,t)$.
The subject of the discussion, which constantly appears in the literature,
is if the predictions of the physical consequences of the theory are
invariant. Two point of view clashed at this point. Some people suppose,
that the longitudial gauge is physically preferred because its
main object, $\Phi(\vec x,t)$, is invariant itself
\cite{bran92}.
The opposite claim is that $\Phi(\vec x,t)$ describes effects in the
fixed frame of reference \cite{unruh,gri94};
in other frames of reference the
observed phenomena may look in another way.

The question under discussion is of principal value now, because
in the coming years
experimental researches of the relic density perturbations spectrum
will be performed with the help of the space apparatus. In fact,
the question is if the theory  in the longitudial gauge can
be used to reconstruct the past of the Universe on the basis
of the observations that will be performed. The Fourier-image
of $\Phi_{\vk}(t)$ is the functional   $\Phi_{\vk}\{H(t)\}$,
and the Hubble function $H(t)$ contains information about the
inflanton field and the inflationary process  itself.  The theoretical
reconstruction of the past is possible only in the case of
$\Phi_{\vk}\{H(t)\}$ is an invariant functional. On the other hand,
in the case of the information contained in $\Phi_{\vk}\{H(t)\}$
essentially depends on the properties of the frame of reference that
is prescribed by the longitudial gauge, the theory of relativistic scalar
potential can not be used to interpret the results of the experiment.

Our claim --- that the theory in the longitudial gauge is noninvariant
--- is made on the basis of the proposed invariant dynamics of the scalar
perturbations.

The main result of our investigation is the strict proof of the
following statement. {\bf The invariant information about the dynamics
of the scalar perturbations is selected from the equations of
the linear  gravitational instability  theory  by the identical
mathematical transformations without making use of any gauge condition
at any stage of the calculations.} Our theory is formulated
as a closed system of equations for the invariant $J_\vk$
of metric perturbations, invariant function $\chi_{\vk {\rm inv}}$
of the perturbations of the inflanton field, its derivative
$\dot{\chi}_{\vk {\rm inv}}$ and energy density perturbations
$\de\ve_{\rm inv}$
(see Eqs. (\ref{inv}), (\ref{chi_inv}), (\ref{e_inv})).
The theory is such that the invariance of the
physical values follows  from their mathematical definitions,
and the invariance of the equations does from the way to obtain them
without any gauge.

For one of the simplest model of inflanton field we perform the numerical
analysis of the evolution of the invariant physical values
and compare the results with those obtained in the longitudial gauge theory.
The general qualitative properties --- the power-law instability of
long--length perturbations and the formation of the Harrison--Zel'dovich
spectrum --- are the same in the two approaches,
but the numerical values are different, the perturbations
in the longitudial gauge theory being several times greater. We find out
that  the reason of the quantitative
differences is the strong perturbations of the proper time in frame of
reference specified by the longitudial gauge.

Our point of view is that the perturbations
of the proper time can not lead to the large scale structure
formation and  thus the invariant theory must be used to interpret
the space experiments.
Among the theories formulated in the fixed gauges,
the synchronous gauge theory  proposed by Lifshitz
\cite{lif} is an adequate one,
because it this gauge the evolution of the background and perturbations
is analyzed in one and the same time. This physical requirement
was formulated in our papers \cite{maro75} devoted to the
description of  the back reaction of the perturbations on the
background.

\section{The equations and the numerical results.}

As it was mentioned in the Introduction the object of our
theory is invariant $J_k$ constructed from the components
of the perturbed metric
\beq
h_i^k(\vk)=\left(
\begin{array}{cc}
\di \Phi_\vk & \di  ik_{\al} a \si_\vk \\
\di  -ik^{\be}\frac{\si_\vk}{a} \quad &
                    \di  \frac13(\mu_\vk+\la_\vk)\de_{\al}^{\be}
       -\frac{k_{\al}k^{\be}}{k^2}\la_\vk
\end{array}  \right),
\quad
J_\vk=\left(\frac{\mu_\vk+\la_\vk}{H} \right)^{\dot{}} - 3\Phi_\vk,
\label{hik_lif}
\eeq
and the invariant functions for the  perturbation of the inflanton field
and energy density
\beq
\begin{array}{c} \di
\chi_{\vk\;{\rm inv}}=\de\phi_\vk-\dot{\phi}\de\tau_\vk(t),
\qquad
\dot{\chi}_{\vk\;{\rm inv}}=\de\dot{\phi}_\vk
-\ddot{\phi}\de\tau_\vk(t)-\frac12\dot{\phi}\Phi_\vk,
\\[3mm] \di
\de\ve_{\rm inv}=\de\ve-\dot{\ve}\de\tau_\vk(t).
\end{array}
\label{4}
\eeq
In (\ref{4})
\[
\de\tau_\vk(t)=\de\int \sqrt{g_{00}} dt=const+\frac12\int\limits_0^t
         \Phi_\vk dt
\]
is the perturbations of the proper  time. In the expressions
for the observed values the function $\de\tau_\vk(t)$ describes
the effects that are related to the noninertial motion of the
frame of reference with respect to the background. So, these effects
are due to the influence of the scalar perturbations on the clock run
in the frame of reference which the scalar perturbations are studied in.

The identical transformations of the equations of the gravitational
instability theory without making use of any gauge lead to the
following equation for the invariant metric function
\beq
\begin{array}{l} \di
\ddot{J}_\vk+
\left( 3H+2\frac{\dot{H}}{H}-\frac{\ddot{H}}{\dot{H}}\right)\dot{J}_\vk
\\[3mm]   \di \hspace*{20mm}
+\left(
\frac{k^2}{a^2}+6\dot{H}+2\frac{\ddot{H}}{H}-2\frac{\dot{H}^2}{H^2}
   -\frac{\stackrel{...}{H}}{\dot{H}}
+\frac{\ddot{H}^2}{\dot{H}^2}-3\frac{H\ddot{H}}{\dot{H}}\right)J_\vk=0
\end{array}
\label{inv}
\eeq
This equation is valid for any potential $U(\vp)$; a fixed potential
makes function  $H(t)$ and its derivatives also fixed.
The relations which connect the invariant characteristics of the
gravitational and inflanton field perturbations are
\beq
\begin{array}{c} \di
\chi_{\vk\;{\rm inv}}=-\frac{1}{3\dot{\vp}}
\left( HJ_\vk+\dot{H}\int\limits_0^t J_\vk dt \right),
\\[5mm]  \di
\dot{\chi}_{\vk\;{\rm inv}}=-\frac{H}{3\dot{\vp}}
\left[ \dot{J}_\vk+\left(2\frac{\dot{H}}{H}-\frac12\frac{\ddot{H}}{\dot{H}}
   \right) J_\vk +\frac{\ddot{H}}{2H}
\int\limits_0^t J_\vk dt \right].
\end{array}
\label{chi_inv}
\eeq
In particular, these equations are used to specify the initial
conditions for invariant and its derivative via those for the
perturbations of inflanton field and its derivative.

For the physical and cosmological applications of the theory
the relative energy density  perturbations  are of particular
interest

\beq
\frac{\de\ve_{\vk\,{\rm inv}}}{\ve} =  -\frac{1}{9H}\left[
\dot{J}_\vk+\left(2\frac{\dot H}{H}-3H-\frac{\ddot H}{\dot H}\right)J_\vk
\right] +\frac{\dot{H}}{3H} \int\limits_{0}^{t} J_\vk dt.
\label{e_inv}
\eeq

We will compare our results with those in the longitudial gauge.
The relativistic scalar potential satisfies the equation
\beq
\ddot{\Phi}_\vk+\left( H-\frac{\ddot{H}}{\dot{H}}\right)\dot{\Phi}_\vk
+\left(\frac{k^2}{a^2}+2\dot{H}-\frac{H\ddot{H}}{\dot{H}}\right)
\Phi_\vk=0
\label{Phi}
\eeq

The perturbations of the inflanton field and energy density
are related to $\Phi_\vk$ as
\beq
\chi_\vk=\frac1{\dot{\vp}}\left(\dot{\Phi}_\vk+H\Phi_\vk\right)
\qquad
\frac{\de\ve_\vk}{\ve}=-\left( \frac{k^2}{3H^2 a^2}\Phi_\vk+1\right)\Phi_\vk
          -\frac{\dot{\Phi}_\vk}{H}
\label{chi_de}
\eeq
The comparison of (\ref{inv}),  (\ref{chi_inv}), (\ref{e_inv}) and
(\ref{Phi}), (\ref{chi_de}) shows that the mathematical structure
of invariant theory and that of the theory in the longitudial gauge
are essentially different. This is  not surprisingly, because the
invariant theory by definition exclude the effects coming from
the perturbations of the proper time.  Contrary, in the frame of reference
specified  by the longitudial gauge, the observed phenomena
strongly depend on the properties of the frame of reference itself,
because the scalar relativistic potential $\Phi_\vk$ describes not only
the physical perturbations but also the perturbations of the proper time.

We perform a formal mathematical analysis and show that (i) the invariant
formulation of the theory is compatible with any gauge;
the set of the gauge only fixes the expressions for
$\chi_\vk-\chi_{\vk\,{\rm inv}}$,
$\dot{\chi}_\vk-\dot{\chi}_{\vk\,{\rm inv}}$,
$\de\ve-\de\ve_{\rm inv}$; (ii) the invariant effect can be
separated from the results obtained in the longitudial gauge theory.
In this gauge both invariant
and noninvariant parts of energy density perturbations are expressed via
$\Phi_\vk$. The invariant part exactly coincide with (\ref{e_inv}),
the function  $J_\vk$, which satisfies equation (\ref{inv}),
being now defined as
$J_\vk{}_{\rm (l.g.)}=-((\Phi_\vk/H)^{\dot{}})+\Phi_\vk)/3$.
The noninvariant part, which is absent, for example,
in synchronous frame of reference,   is the longitudial gauge
is
\[
\de\ve_\vk{}_{\rm noninv}=\dot{\ve}\de\tau_\vk=
H\dot{H}\left( const+
\frac1{a} \int\limits_0^t J_\vk{}_{\rm (l.g.)} a dt -
\int\limits_0^t J_\vk{}_{\rm(l.g.)}  dt  \right)
\]

Unfortunately, such separation of effects can not be performed in
the longitudial gauge  theory as it is; it is necessary to know
beforehand, what effects are invariant. We think, that is the reason
why the physical effects were not separated from the coordinate ones
in previous works and so the total effect were taken into account in
physical and cosmological applications of the theory.

We perform the numerical comparison of the invariant effects  with
the total effects obtained in the longitudial gauge.
The comparison is made in the framework of the simplest model
of inflation with the inflanton potential $U(\vp)=m^2\vp^2/2$,
where $m=0.15$ in the system of units $\hbar=c=8\pi G=1$.
In the model under consideration the slow-rolling regime holds
form $t=0$ till $t_{\rm inf}=125$, e-folding parameter being equal to 70.

On Figs.1,2 the  relative values of inflanton field perturbations
and energy density perturbations are shown in the long--length--wave
limit $k \ll a_0 H_0$.
The normalization is made with respect to the corresponding values in the
initial moment of time.  The effect of the power-low instability
holds both in the invariant theory and in the longitudial
gauge theory.  One can see from Figs. 1a, 2a, the differences in the grows
of the perturbations take place from the very beginning of the inflation.
By the end of the inflation, as one can see form Figs. 1b, 2b,
the differences  in the predictions of the theories achieve 4-5 times,
the instability being stronger in the longitudial
gauge theory. The physical reason of such differences is that
the measurements of the background and the perturbations are
nonsynchronized in the frame of reference specified by the
longitudial gauge. The background physical parameters are measured
in some later moment  of time, when their values
have decreased during the inflation. The result that the longitudial
gauge theory yields the overestimated values for the long-length--wave
perturbations was also obtained in \cite{gri94} in the framework of
the theory in the synchronous gauge.

The results of the invariant theory answer the question about
the  stability of inflation in the slow-rolling regime.
For the simplest model discussed here, Figs. 1b, 2b show
that the relative perturbation  of the inflanton field increases
by the end of inflation in 25 times, the relative perturbation
of the energy density  does in 15 times.  It is known,
the end of the inflation is
the most important epoch for the formation of the
observed properties of the Universe, so we need to discuss the influence
of the perturbations on this regime.
The initial long--length--wave  perturbations
of the inflanton field $\chi(0)/\vp(0)$, generally speaking, is
a random value the nature of which must be clarified in quantum
geometrodynamics; the value  $\chi(0)/\vp(0)\sim 0.1$  seems to be
physically reasonable.  In this case the long--length--wave
perturbations generate the additional stochastic  effects comparable with
the effects taken into account in the background solution. In particular,
the evolution of the system in the end of the inflation must be strongly
influenced by the back reaction  of the perturbations on the background.
The invariant theory of the back reaction of the scalar perturbations
on the background is the subject of another paper \cite{our}.
We would like to emphasize, such back reaction  plays an
important role in forming the intermediate stage, when  the
inflationary  epoch changes  to  the  Friedman one.
It is this intermediate stage that the  final properties of the
Universe which are the subject of coming cosmic experiments are formed.

The next problem is the theory of quantum fluctuations of the
inflanton field, which are
short--length--wave ones in the beginning of the inflation.
As it is known, the theory of these fluctuations is, in fact,
the theory of relic density perturbations, the evolution of which
have lead to the formation of the  large scale structure of the
Universe. It is necessary to differ three stages of relic perturbations
formation: (i) the stage of quantum oscillations, on which
$k \gg aH$; (ii) the state of aperiodic increase from the
moment of the oscillation termination till the end of inflation;
(iii) the stage of spectrum reconstruction
(metric preheating \cite{bas}) on which the final formation of the
relic perturbation spectrum takes place. The quantum theory of fluctuations
is able to describe the evolution of fluctuations on all the three
stages. On the  first and the second stages the analysis can be
performed on the given background, which correspond to the
slow-rolling regime. On the third stage, as it was mentioned above,
the back reaction of the perturbations must be taken into account.

Below we'll describe the results of investigation on the first and
the second stages.  Generally speaking, there exists a method of numerical
estimations of relic perturbation level and spectrum at
the end of inflation, which do not use the gravitational instability
theory \cite{lyth99}. Notice at once, the results of the proposed
invariant theory of quantum fluctuation are in quantitative agreement
with this estimations. However, the invariant perturbation
theory is  more regular method, because it allow to describe
successively all the three stages of spectrum formation.

Now let us pass to the quantum perturbation theory  as it is.
First of all, notice,
there is a formal problem in the theory: what function is the object
of quantization. The pose of this problem is induced by our understanding
that the coordinate effects that can be eliminated by the appropriate
choice of a classical frame of reference must not to be subject to
quantization. In the longitudial gauge theory the discussion of this
problem is again reduced to the question whether the description
in terms of the relativistic potential is invariant  or not.
One of the achievements of our invariant theory is that this question
is uniquely solved. Only the invariant metric function which
is uniquely connected with the invariant characteristics of  the
inflanton field is the subject of quantization.  For the
short--length--wave
fluctuations $k/a_0 H_0 \gg 1$ the normalization of quantum
operator is easy to find:
\beq
\hat{J}_\vk\approx \frac{3\dot{\vp}}{H}\hat\Psi_\vk,
\quad
\hat{\Psi}_\vk\approx \frac{1}{\sqrt{2k}a}
\left[ \hat{c}_\vk \exp\int\limits_0^t \frac{k\, dt}{a}
+\hat{c}^+_{-\vk} \exp \left(-\int\limits_0^t
                        \frac{k\, dt}{a}\right) \right],
\quad \frac{ka}{H}\gg 1
\label{quant}
\eeq
where $\hat{c}_\vk$ and $\hat{c}^+_{-\vk}$ are annihilation and creation
operators in quantum field theory.
Notice, expressions (\ref{quant}) are used only to specify the
initial conditions; the quantum dynamics itself is described by
the exact operator equation (\ref{inv}).
 The subject of calculations
is the value of energy density fluctuations averaged over
the Heisenberg vacuum specified in the beginning of the inflation.
\beq
\left| \frac{\de\ve_{\vk\, {\rm inv}}}{\ve} \right|
\equiv \sqrt{ \bl 0| \left(
\frac{\de\ve_{\vk\, {\rm inv}}}{\ve}
                     \right)^2 | 0 \br}
\label{9}
\eeq
In order to calculate the value (\ref{9}) one can solve
equation (\ref{inv}) under $|c_\vk|=|c_{-\vk}^+|=1/2$.
The averaging over the phases of the complex
numbers  $c_\vk$, $c_{-\vk}$ and taking the absolute value
corresponds to the procedure of quantum averaging.

In the longitudial gauge theory we can set the subject
of quantization only by the formal agreement without strict
proof. The quantum relativistic potential is
\beq
\dot{\Phi}_\vk \approx \dot{\vp}\hat{\Psi}_\vk
\approx -\frac{i\dot{\vp}}{\sqrt{2}k^{3/2}}
\left[ \hat{c}_\vk \exp\int\limits_0^t \frac{k\, dt}{a}
- \hat{c}^+_{-\vk} \exp \left(-\int\limits_0^t
                        \frac{k\, dt}{a}\right) \right]
\eeq

We'll pass to the comparison of the results.
For quantum fluctuations of the energy density
there is a region of oscillations, which is the more wide, the less
the initial length of the wave is.
In the model under consideration, $k\sim 10^6$ corresponds to the
large scale structure of the modern Universe. In order to
demonstrate the physical and mathematical contents of the theory,
we use $k=50$, for which the detailed  computer calculations
are easy to done.  The evolution of quantum fluctuations on the
region of oscillations is shown on Fig. 3a for the invariant theory and
Fig. 3b for the longitudial gauge theory. One can see, there is no
difference in the predictions of the two theories in this region.
This conclusion is physically clear, because  the effect of
clocks nonsynchronization in the longitudial gauge theory is
very small in comparison with quick change of the amplitude
of quantum fluctuations, which are dictated by the proper dynamics
of this fluctuations. However, the quantum fluctuations
inevitable goes to the  aperiodic evolution stage, which is shown
on Fig. 3c. Here the strong quantitative difference between the
two theories is evinced.

The theoretical predictions  for the parameter of the
Harrison--Zel'dovich spectrum are of the most interest.
This parameter is defined as follows
\beq
\Delta= \left(
 \bl 0| \left(
\frac{\de\ve_{\vk\, {\rm inv}}}{\ve}
      \right)^2 | 0 \br \frac{k^3}{2\pi^2} \right)^{1/2}.
\eeq
The results of calculation of $\Delta$ without the averaging over
the initial  phases of quantum fluctuations are shown on Fig. 4.
As it was expected, in the both theories we have a flat spectrum.
However, the numerical values in the longitudial gauge theory are
strongly high than those in the invariant theory.
As it was noticed  earlier, the difference is generated by
the effect of the clock unsynchronization on the stage of the
aperiodic evolution.

So, before the interpretation of the results of the cosmic experiments,
we must clarify, what information is in the results of the measurements.
If we are sure that we are able to separate the invariant information,
then the invariant theory  is needed for its theoretical interpretation.

\section*{Appendix.}

In this appendix we'll show how the invariant equation (\ref{inv})
can be obtained.

The perturbed metric (\ref{hik_lif}) is substituted to the perturbed
Einstein equations $\de R_i^k-\frac12 \de_i^k R=\de T_i^k$,
where $\de T_i^k$ is energy--momentum tensor of inflanton field,
the potential being arbitrary. From $({}_0^0)$   and $({}_\al^0)$
Einstein's equations the perturbations of inflanton field are expressed
via perturbations of metric, this allows to write the $({}_\al^\be)$
Einstein equation down in the form of the two  following
equations for three variables
\beq
\ddot{N}_\vk+3H\dot{N}_\vk-\frac{k^2}{3a^2}(N_\vk+3\Phi_\vk)-\dot{L}_\vk
-3HL_\vk=0
\label{cosm1}
\eeq
\beq
L_\vk=-\frac{1}{3H}\left[
\ddot{N}_\vk-\left(3H+\frac{\ddot{H}}{\dot H} \right)\dot{N}_\vk
+\frac{k^2}{a^2}N_\vk-3H\dot{\Phi}_\vk
-3\left(2\dot{H}-\frac{H \ddot{H}}{\dot{H}} \right)\Phi_\vk\right]
=0
\label{cosm2}
\eeq
Here
\[
N_\vk=\mu_\vk+\la_\vk, \quad
\quad L_\vk=\dot{\mu}_\vk+2 ik\si_\vk.
\]

Equations (\ref{cosm1}), (\ref{cosm2}) show that all  gauges  are to
be presented in the form of the additional constraints for
$N_\vk$, $L_\vk$, $\Phi_\vk$. In particular, the synchronous gauge
is $\Phi_\vk=0$, the longitudial gauge is $L_\vk=\dot{N}_\vk$.
It is of principal importance, however,
equations (\ref{cosm1}), (\ref{cosm2}) allows to separate
the invariant information without making use of any gauge.
Substituting (\ref{cosm2}) to (\ref{cosm1}) lead to the equation in which
the variables $N_\vk$, $\Phi_\vk$ are present only in the invariant
combination $J_\vk=(N_\vk/H)^{\dot{}}-3\Phi_\vk$,
 which is a linear superposition
of two Bardeen's invariant. The equation obtained is
the desired equation of invariant dynamics (\ref{inv})
that was written in Section 2.

\begin{figure}[htb]
\centering{ \mbox{\epsfig{figure=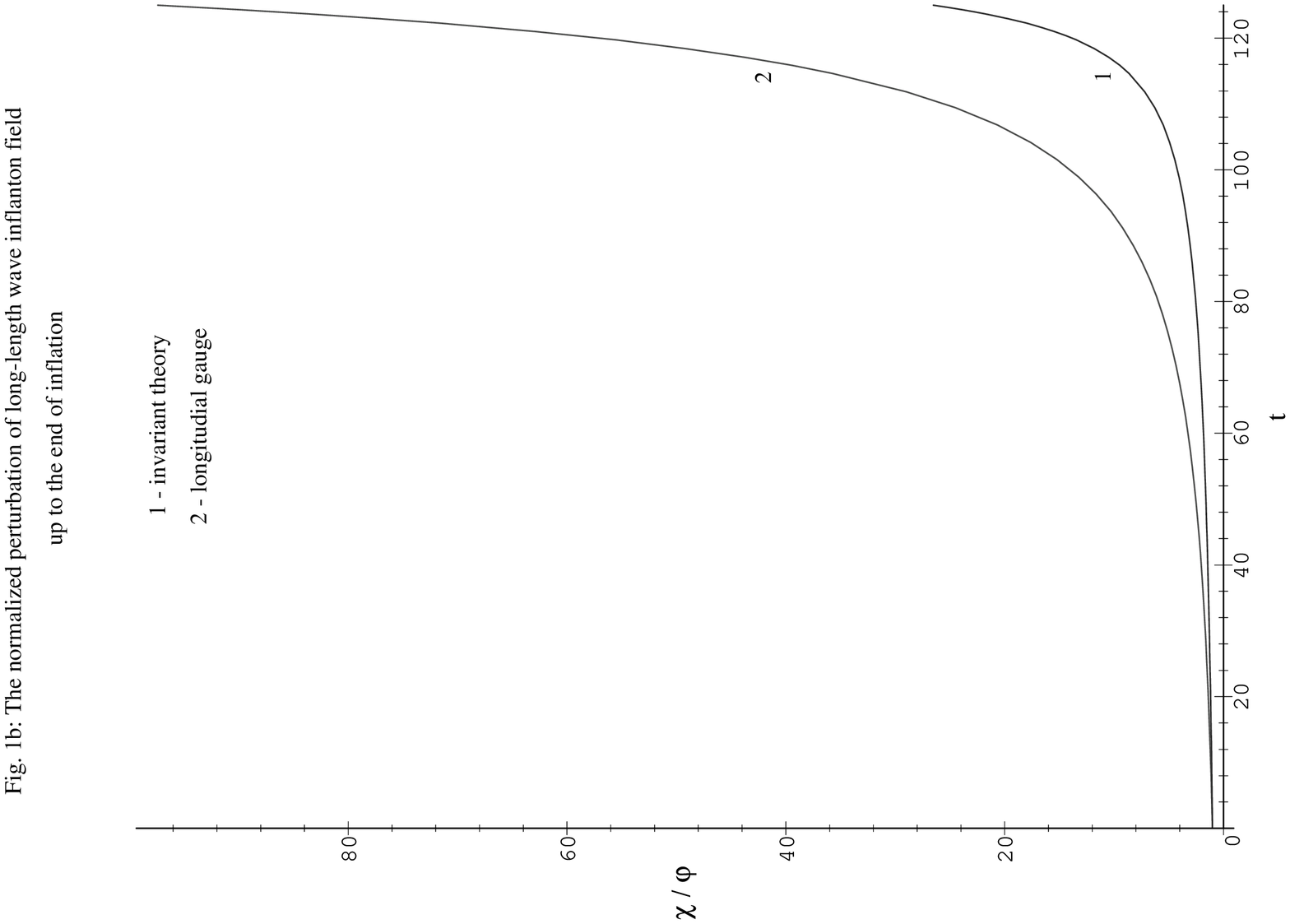,width=14.5cm,angle=90}}
\mbox{\epsfig{figure=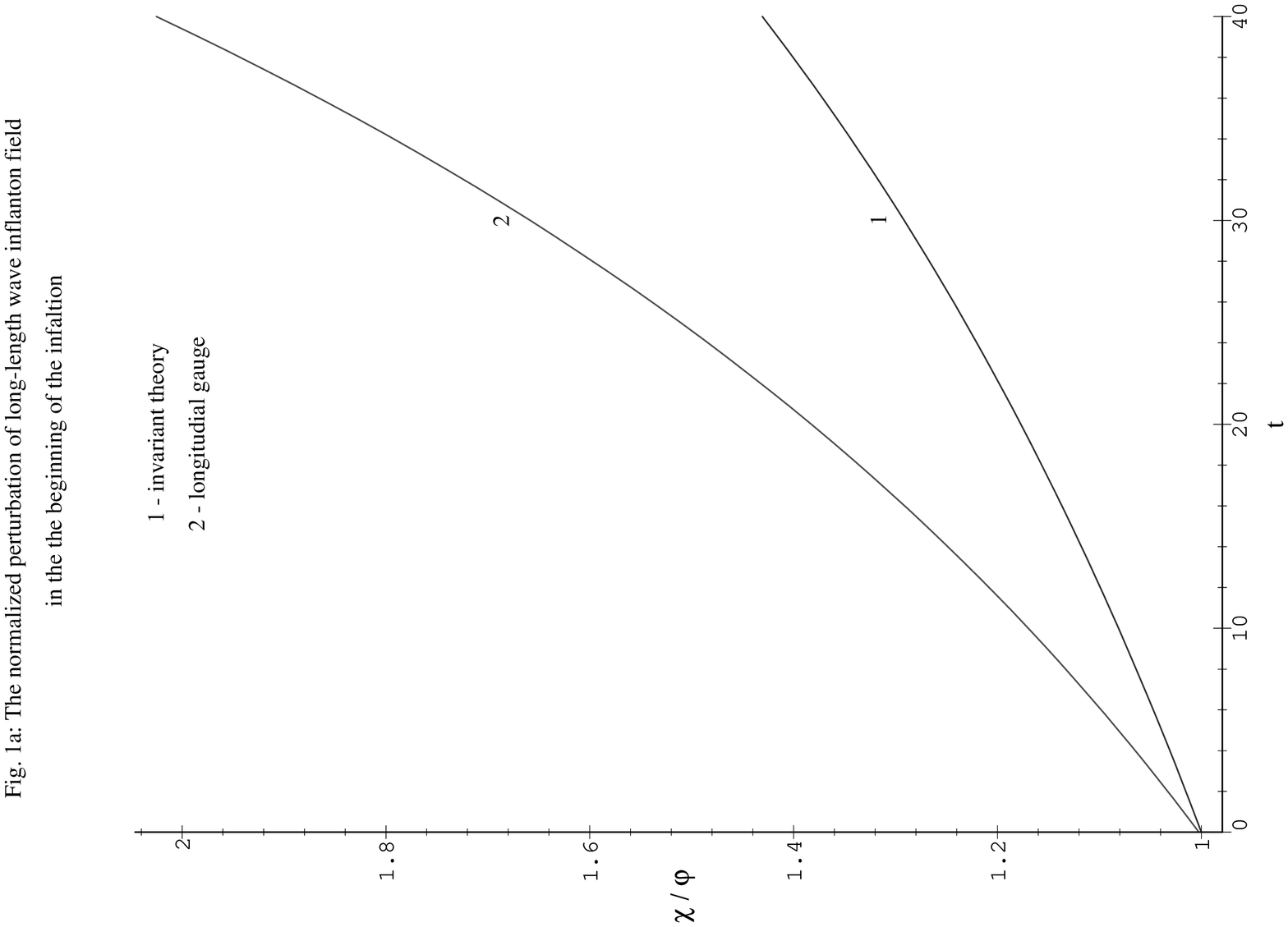,width=14.5cm,angle=90}}}
\end{figure}
\begin{figure}[htb]
\centering{ \mbox{\epsfig{figure=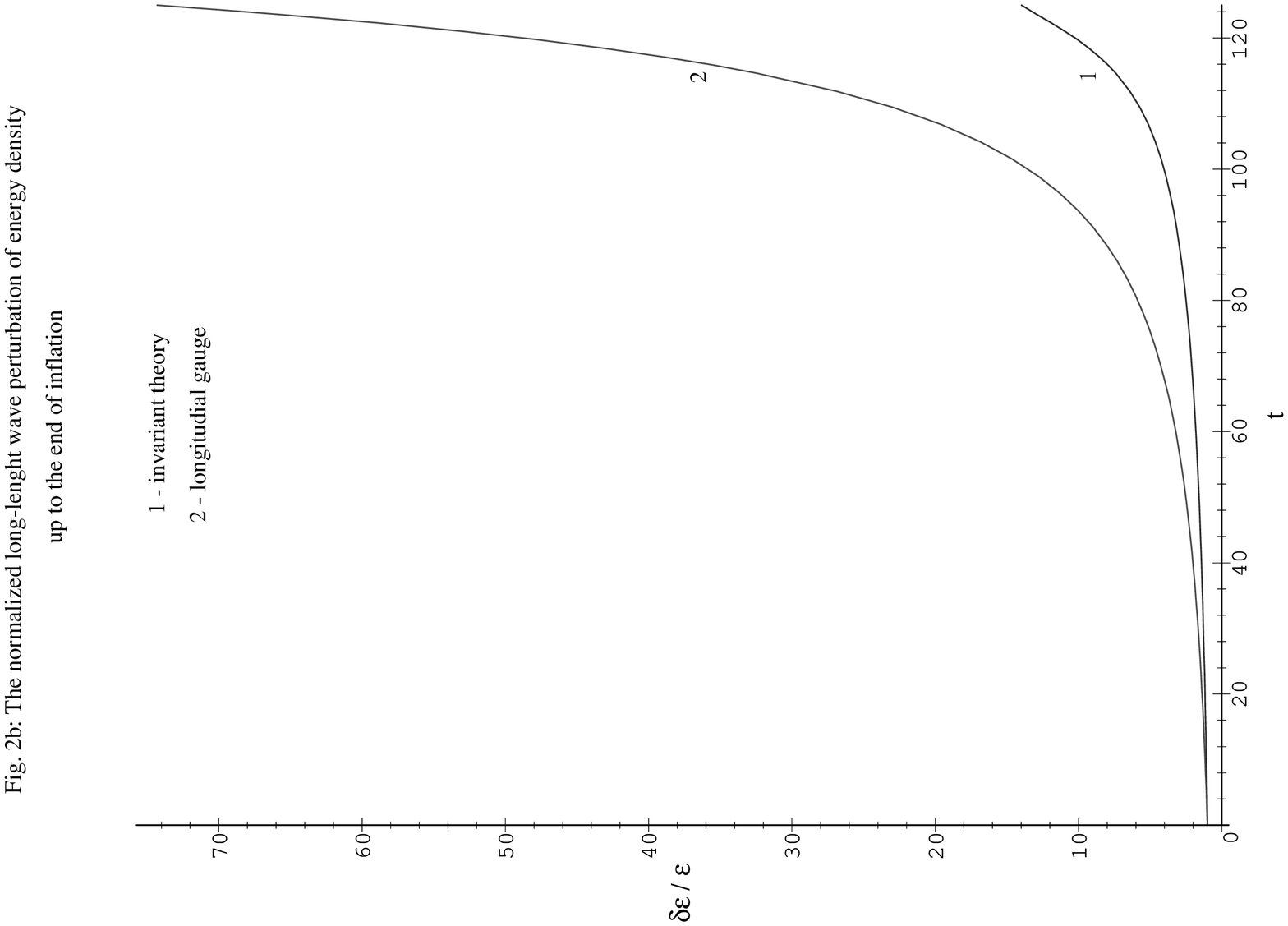,width=14.5cm,angle=90}}
 \mbox{\epsfig{figure=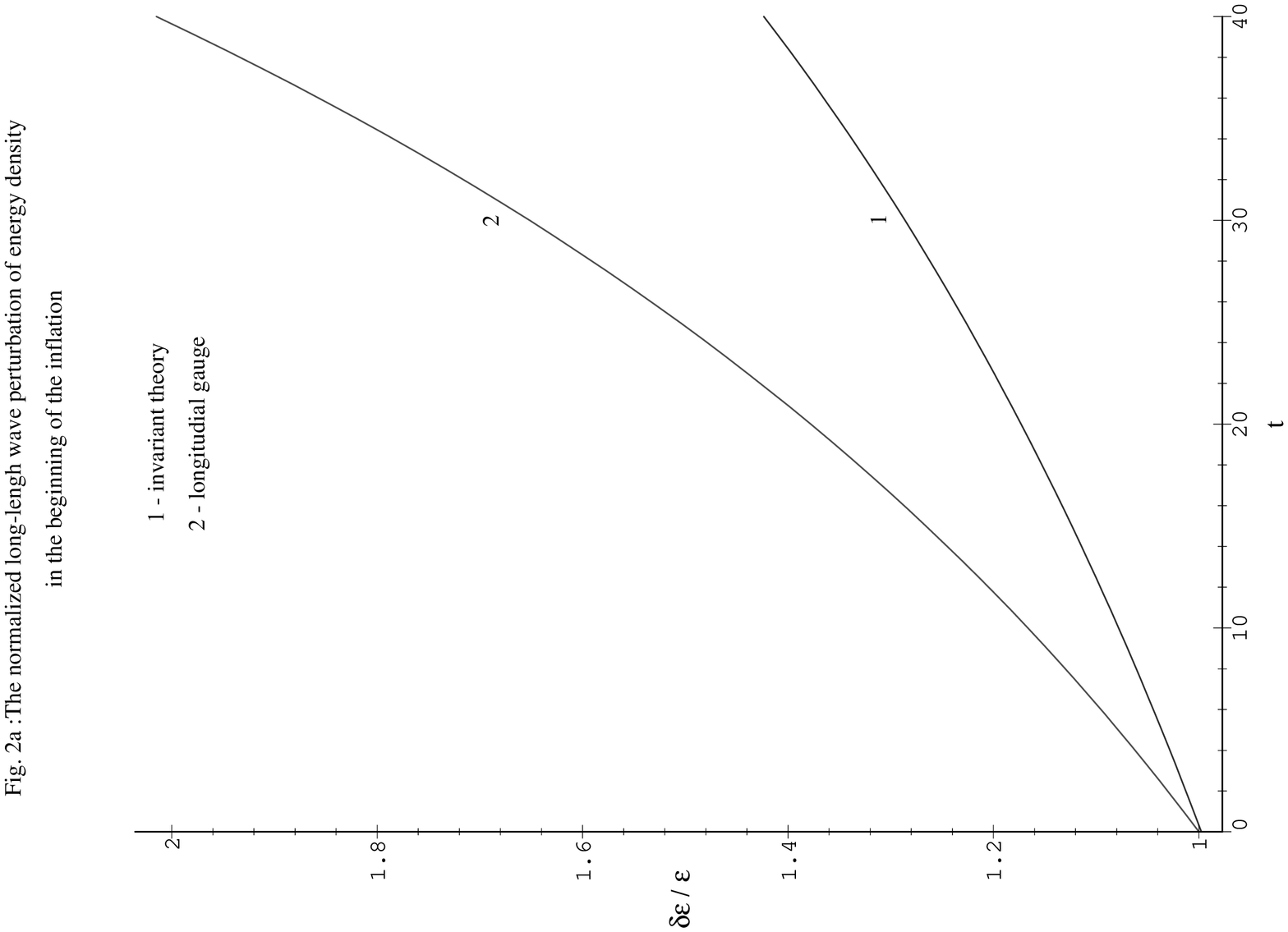,width=14.5cm,angle=90}}}
\end{figure}
\begin{figure}[htb]
\centering{ \mbox{\epsfig{figure=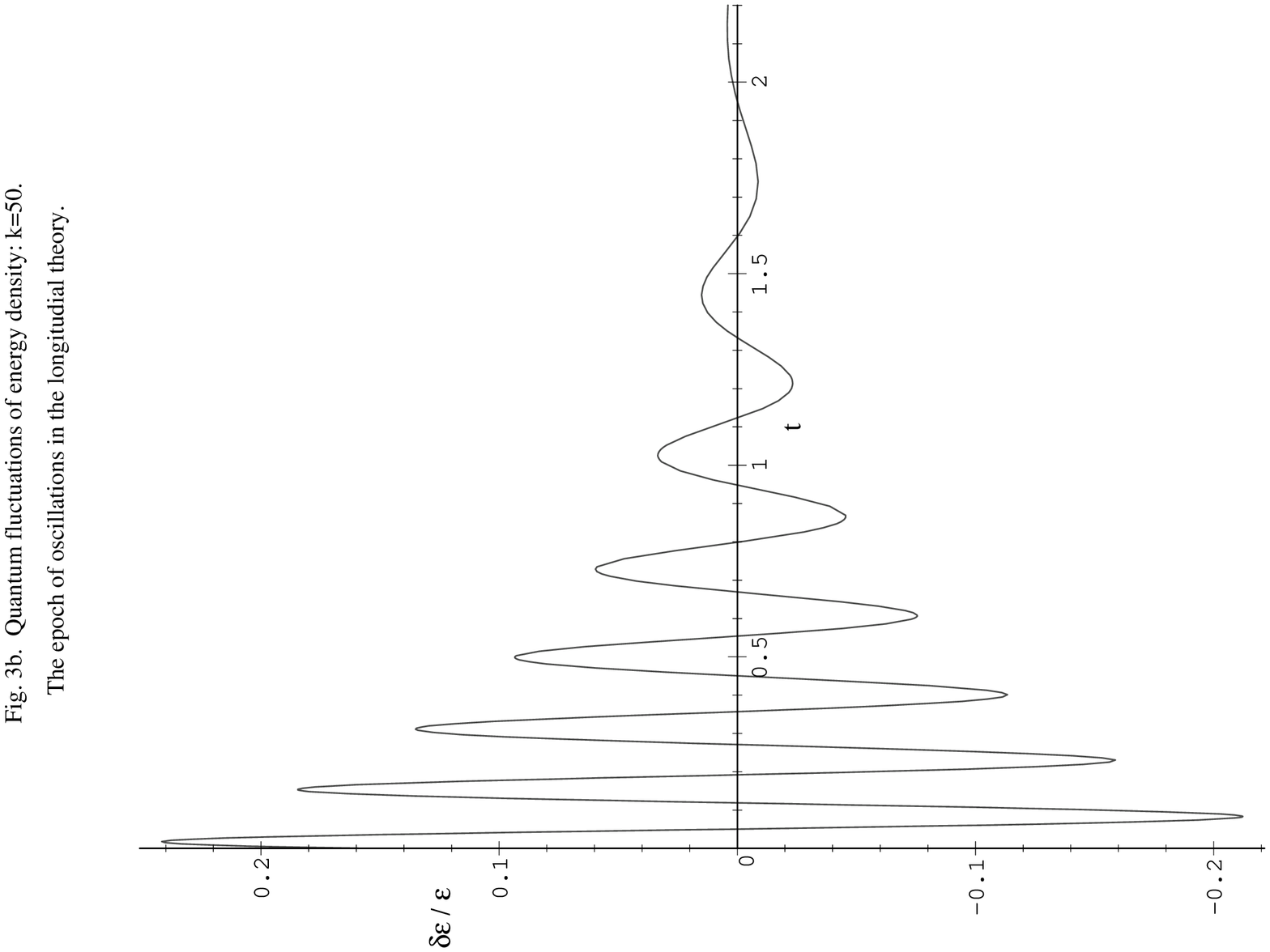,width=14.5cm,angle=90}}
 \mbox{\epsfig{figure=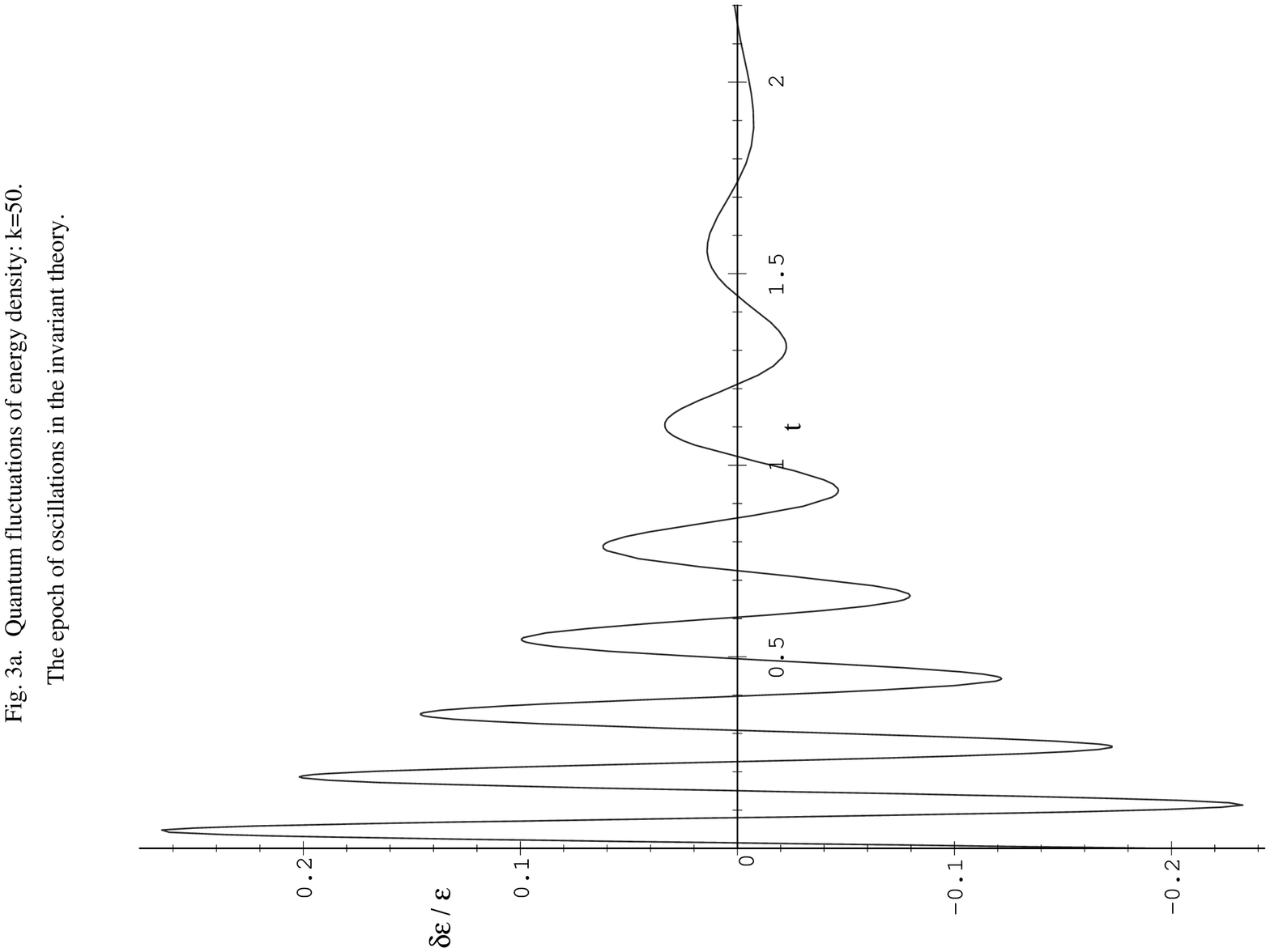,width=14.5cm,angle=90}}}
\end{figure}
\begin{figure}[htb]
\centering{ \mbox{\epsfig{figure=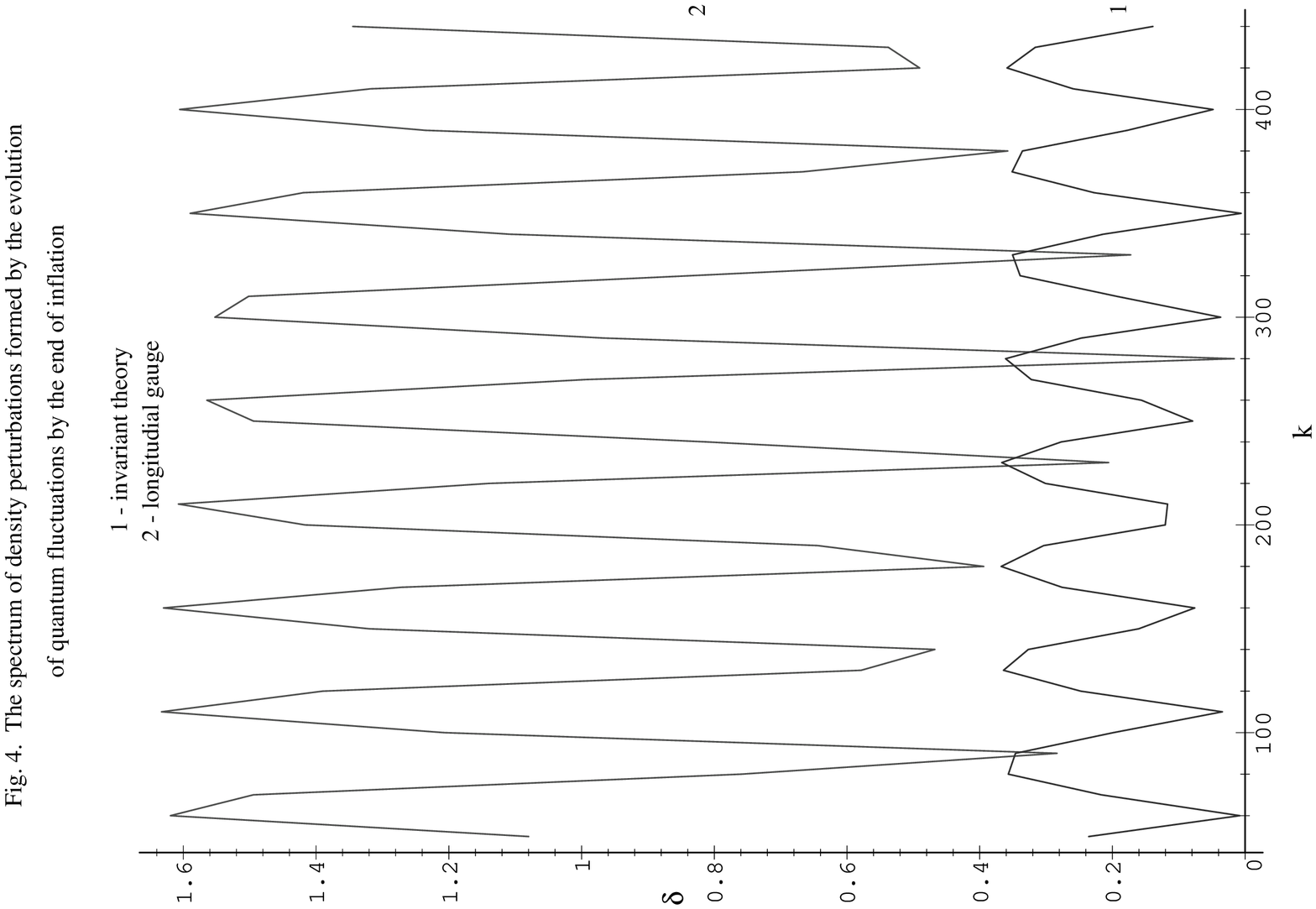,width=14.5cm,angle=90}}
 \mbox{\epsfig{figure=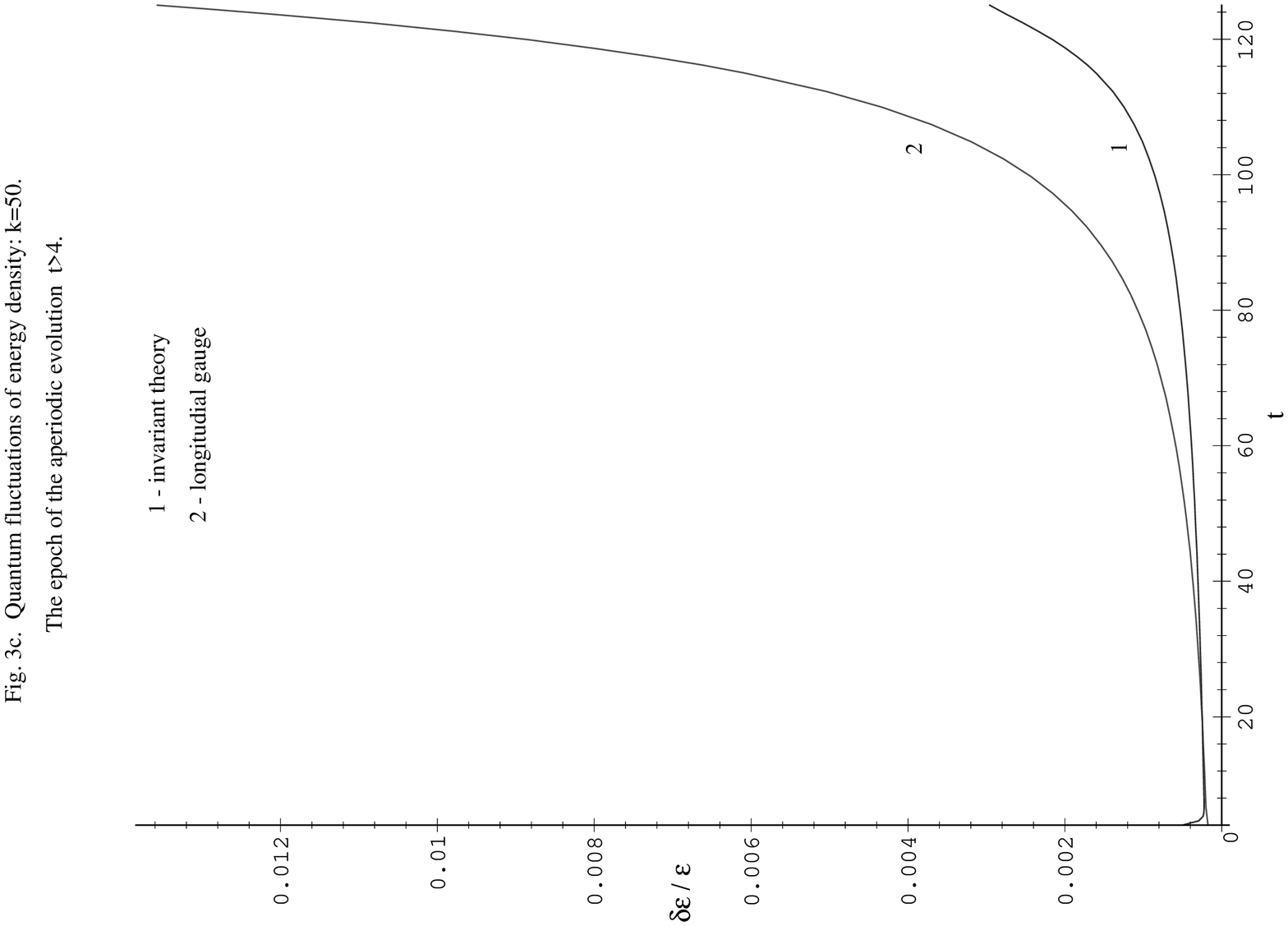,width=14.5cm,angle=90}}}
\end{figure}

\end{document}